\documentclass[10pt,preprint2]{aastex}
\received{2009 April~28}
\revised{2009 August~18}
\accepted{2009 October~19}
\paperid{316756}
\shorttitle{Planetary Magnetospheric Emission Survey}
\shortauthors{Lazio et al.}

\newcommand{\mjybm}{\mbox{mJy~beam${}^{-1}$}}

\begin{document}
\title{A Blind Search for Magnetospheric Emissions from Planetary
	Companions to Nearby Solar-type Stars}

\author{T.~Joseph~W.~Lazio\altaffilmark{1}}
\affil{Naval Research Laboratory, 4555 Overlook Avenue \hbox{SW},
	Washington, DC 20375-5351}
\altaffiltext{1}{NASA Lunar Science Institute, NASA Ames Research Center, Moffett Field, CA}

\email{Joseph.Lazio@nrl.navy.mil}

\and

\author{%
S.~Carmichael,\altaffilmark{2}
J.~Clark,\altaffilmark{3}
E.~Elkins,\altaffilmark{4}
P.~Gudmundsen,\altaffilmark{5}
Z.~Mott,\altaffilmark{6}
M.~Szwajkowski,\altaffilmark{7}
L.~A.~Hennig}
\affil{Thomas Jefferson High School for Science and Technology, 
	6650 Braddock Road, Alexandria, VA  22312}

\altaffiltext{2}{Current address: George Mason University}
\altaffiltext{3}{Current address: University of Notre Dame}
\altaffiltext{4}{Current address: University of Virginia}
\altaffiltext{5}{Current address: Princeton University}
\altaffiltext{6}{Current address: College of William \& Mary}
\altaffiltext{7}{Current address: Northwestern University}

\begin{abstract}
This paper reports a blind search for magnetospheric emissions from
planets around nearby stars.  Young stars are likely to have much
stronger stellar winds than the Sun, and because planetary
magnetospheric emissions are powered by stellar winds, stronger
stellar winds may enhance the radio luminosity of any orbiting
planets.  Using various stellar catalogs, we selected nearby stars
($\lesssim 30$~pc) with relatively young age estimates ($< 3$~Gyr).
We constructed different samples from the stellar catalogs, finding
between~100 and several hundred stars.  We stacked images from the
74-MHz (4-m wavelength) VLA Low-frequency Sky Survey (VLSS), obtaining
3$\sigma$ limits on planetary emission in the stacked images of
between~10 and~33~mJy.  These flux density limits correspond to
average planetary luminosities less than 5--10 $\times
10^{23}$~erg~s${}^{-1}$.  Using recent models for the scaling of
stellar wind velocity, density, and magnetic field with stellar age,
we estimate scaling factors for the strength of stellar winds,
relative to the Sun, in our samples.  The typical kinetic energy
carried by the stellar winds in our samples is 15--50 times larger
than that of the Sun, and the typical magnetic energy is 5--10 times
larger.  If we assume that every star is orbited by a Jupiter-like
planet with a luminosity larger than that of the Jovian decametric
radiation by the above factors, our limits on planetary luminosities
from the stacking analysis are likely to be a factor of~10--100 above what
would be required to detect the planets in a statistical sense.
Similar statistical analyses with observations by future instruments,
such as the Low Frequency Array (LOFAR) and the Long Wavelength Array
(LWA), offer the promise of improvements by factors of~10--100.
\end{abstract}

\keywords{planetary systems}

\section{Introduction}\label{sec:intro}

In searches for extrasolar planets, the precision to which a planetary
signal can be extracted from the data can depend in part on the
properties of the host star.  For instance, in radial velocity
surveys, one of the limiting factors in the velocity precision is
intrinsic stellar ``jitter,'' caused by starspots or other surface
inhomogeneities.  Such stellar jitter is well known to be correlated
with stellar activity, the level of which declines in age
\citep{bmwmdv96,sd97}.  Radial velocity surveys tend to select stars
that are chromospherically quiet \citep{sbm98,cbmvwf08}, which is
likely to introduce a bias toward older stars.  Further, the link
between chromospheric activity and age means that distinguishing
planetary transits from stellar surface features will probably be
easier for older, less active stars \citep[but see][]{j02}.
Consequently, there is a selection bias \emph{against} planets around
younger (``adolescent'') stars.

All of the solar system's ``magnetic'' planets (\objectname[]{Earth},
\objectname[]{Jupiter}, \objectname[]{Saturn}, \objectname[]{Uranus},
and \objectname[]{Neptune}) generate planetary-scale magnetic fields
as the result of internal dynamo currents within the planet.  The
solar wind incident on these planetary magnetospheres is an energy
source to the planetary magnetospheres, and the magnetosphere-solar wind
interaction produces energetic (keV) electrons that then propagate
along magnetic field lines into auroral regions, where electron
cyclotron masers are produced.

Specific details of the cyclotron maser emission vary from planet to
planet, depending upon such secondary effects as the planet's magnetic
field topology.  Nonetheless, applicable to all of the magnetic
planets is a macroscopic relation relating the incident solar wind
power $P_{\mathrm{sw}}$, the planet's magnetic field strength, and the
median radio luminosity~$L_{\mathrm{rad}}$.  Various investigators
\citep{db84,dk84,dr85,bgd86,r87,d88,mg88} find
\begin{equation}
L_{\mathrm{rad}} = \epsilon P_{\mathrm{sw}}^x,
\label{eqn:law}
\end{equation}
with $\epsilon$ the efficiency at which the solar wind power is
converted to radio luminosity, and $x \approx 1$.  The value for
$\epsilon$ depends on whether one considers the magnetic energy or
kinetic energy, respectively, carried by the stellar wind.  The strong
solar wind dependence is reflected in the fact that the luminosity of
the Earth is larger than that of either Uranus or Neptune, even though
their magnetic field are 10--50 times stronger than that of the Earth.

The magnetospheric emissions from solar system planets and the
discovery of extrasolar planets has motivated both theoretical
\citep{zqr+97,fdz99,ztrr01,flzbdr04,lfdghjh04,s04,gmmr05,z06,z07,gzs07,gpkmmr07}
and observational work
\citep{yse77,wdb86,bdl00,lfdghjh04,rzr04,gs07,lf08,scgjlb09} on
magnetospheric emissions from extrasolar planets, including some
before the confirmed discovery of any extrasolar planets.

Implicit in many of the early predictions for planetary magnetospheric
emissions was that the stellar winds of other stars are comparable to
the solar wind.  Yet, from measurements of the sizes of astropauses
(i.e., the boundary between the stellar wind and the local
interstellar medium), \cite{wmzl02,wmzlr05} find the mass loss rate as
a function of age, $\dot M \propto t^x$, with $x \approx -2$, a
dependence probably linked to the decrease in surface magnetic
activity with stellar age.  Thus, the stellar wind around a 1~Gyr old
star may be 25 times as intense as the current solar wind (from a
4.5~Gyr old star).

As a specific illustration of the possible effect of considering
younger stars with more intense stellar winds, \cite{fdz99} and
\cite{lfdghjh04} predicted that the magnetospheric emission from the
planet orbiting $\tau$~Boo would be of order 1--3~mJy at frequencies
around~30~MHz (10~m wavelength), {assuming that the stellar wind of
the star was comparable to that of the Sun}.  \cite{s04} and
\cite{gzs07}, taking into account the likely stellar wind strength of
$\tau$~Boo, predicted that its emission would be at a level of order
100--300~mJy.  The former prediction (1--3~mJy) is below the
sensitivity of current instrumentation; the latter is not.

This paper reports a blind search for magnetospheric emissions from
nearby ``adolescent'' stars.  In \S\ref{sec:catalog} we describe how
we selected stars from existing catalogs of nearby stars, in
\S\ref{sec:radio} we present the results from our stacking analyses
and as well as considering the stars individually, and in
\S\ref{sec:conclude} we present our conclusions.

\section{Stellar Catalog Assembly}\label{sec:catalog}

Three catalogs form the basis for our identification of nearby
``adolescent'' stars.
\begin{enumerate}
\item%
The NStars program is part of the \textit{Space Interferometry
Mission} Preparatory Science Program.  \cite{gcg+03} and \cite{gcg+06}
obtained spectroscopic observations of the 3600 main-sequence and
giant stars within~40~pc of the Sun with spectral types earlier than
M0.

\item%
The Spectroscopic Properties of Cool Stars (SPOCS) is a compilation of
stars forming the basis of radial velocity planetary searches.  From
the series of observations in the SPOCS program, \cite{tfsrfv07}
estimated various physical parameters for the stars.

\item 
The Geneva-Copenhagen survey of the solar neighborhood (GCS) is an
effort to assemble a complete and consistent set of observational and
physical parameters for nearby F and G dwarfs \citep{hna09}.
\end{enumerate}

We applied four selection criteria---spectral type, distance, age, and
declination---to the published catalogs to form samples for further
study.  For spectral type, we selected main-sequence \hbox{F},
\hbox{G}, or~\hbox{K} dwarfs.  The electron cyclotron maser---the same
process by which planets generate radio emission---has been detected
from some stars.  Typical detections are at frequencies of order
5000~MHz, from which (lower) coronal magnetic field strengths of order
300~G or higher are inferred \citep{hadblg08}.  For the electron maser
emission process, the quantity $B/\nu$ is approximately constant,
where $B$ is the magnetic field in the emitting region.  At
frequencies $\nu \approx 100$~MHz, we would expect magnetic field
strengths of order 30~\hbox{G}; lower frequencies would result in even
lower field strengths.  A field strength of~30~G is typical of that
inferred for very late M dwarfs and early L brown dwarfs \citep{b02}
and within a factor of a few of the field strength of Jupiter at the
cloud tops (4~G).  Moreover, while the Sun generates intense emission
at these frequencies, notably Type~III and~IV radio bursts, even the
strongest such radio bursts are far too faint to be detected over
interstellar distances.  \cite{g86} has considered the detection of
solar-type stars at low radio frequencies and finds that current
detection thresholds are at least a factor of~$10^2$ above what is
required.  If the targeted star is an \hbox{F}, \hbox{G}, or K~dwarf,
the presence of low radio frequency emission would be a strong
indication that it is orbited by a sub-stellar companion.

Two of the catalogs, NStars and \hbox{SPOCS}, contain only relatively
nearby stars, with distances less than 40~pc and 130~pc, respectively;
for the SPOCS catalog, fully two-thirds of the stars are
within 40~pc.  The GCS catalog extends to much larger distances,
approximately 1000~pc.  We impose a distance constaint for the
following reason.  

\setcounter{footnote}{7}

In the general stacking analysis, as one stacks images from
increasingly distant sources, the signal-to-noise ratio in a given
image is decreasing, but is compensated by the increasing number of
sources to stack.  Considering stars to lie in shells at distances~$D$
with thickness~$\Delta D$, if all stars host planets, then the flux
densities of planets at larger distances will be smaller by a
factor~$D^2$ which will be exactly balanced by the increase in volume
provided by going to larger distances, assuming a uniform distribution
of stars.  However, it is not yet known whether all stars host
planets, and current limits are that the fraction of stars with
planets is $f_p \approx 0.19$, for planets with semi-major axes less
than about~20~AU\footnote{
As magnetospheric emissions are
powered by the stellar wind, for the purposes of this analysis, the
presence of planets more distant from their host star are not likely
to be relevant.}
\citep{mbv+08}.  Consequently, as one stacks images, an increasing
fraction of the total number of images contain only noise so that the
effective ``signal-to-noise'' ratio is decreasing.  For the GSC
catalog, we impose a distance constraint of~40~pc, for consistency
with the other catalogs.

Two of the catalogs (SPOCS and GCS) report the ages of the stars,
which we adopt.  For both catalogs, not only are ages reported but
also an estimated uncertainty.  Given that stellar ages often have
large uncertainties associated with them, in order not to exclude
stars that might be younger than their nominal ages, we construct two
samples from both catalogs.  One sample requires that the reported age
of the star be less than 3~Gyr (samples that we denote SPOCS-age and
GCS-age), while the other sample requires that the age taking its
uncertainty account is less than 3~Gyr (samples that we denote
SPOCS-eage and GCS-eage).  The NStars catalog does not report the age
of the star, but does report chromospheric flux in the \ion{Ca}{2}~H
and~K lines ($\log R_{HK}^\prime$).  We convert these to estimated
ages following the work of \cite{hsdb96}.

The final selection criterion, declination, is that we will focus on a
northern hemisphere survey for which the effective declination limit
is $-25\arcdeg$ (\S\ref{sec:stack}).  More generally, we are not aware
of an all-sky catalog, at a common frequency and approximately
constant rms image noise level, on which to perform a similar analysis.
Table~\ref{tab:catalog} summarizes various properties of
the sub-samples from the three catalogs.

\begin{deluxetable}{lcccccc}
\tablewidth{0pc}
\tabletypesize{\footnotesize}
\tablecaption{Stellar Catalog Data\label{tab:catalog}}
\tablehead{%
 \colhead{Catalog} & \colhead{Total Number} & \colhead{Magnetospheric} 
	& \colhead{Median} & \colhead{Weighted Average}
	& \colhead{Stacked Image} & \\
 \colhead{Name}    & \colhead{in Catalog}   & \colhead{Subset}
	& \colhead{Age}     & \colhead{Distance} 
	& \colhead{Noise Level}   & \colhead{Reference} \\
                   &                        &
        & \colhead{(Gyr)}   & \colhead{(pc)}
	& \colhead{(\mjybm)}
}
\startdata
NStars     & 664$+$1676 & 252$+$249 & 1.3 & 24.4 & 5.7 & 1,2 \\
SPOCS-age  & 1074       & 110       & 1.9 & 19.1 & 11  & 3 \\
SPOCS-eage & 1074       & 176       & 1.4 & 14.3 & 9.3 & 3 \\
GCS-age    & 16682      & 355       & 1.6 & 21.4 & 6.0 & 4 \\
GCS-Eage   & 16682      & 656       & 0.7 & 23.0 & 4.8 & 4
\enddata
\tablecomments{The NStars catalog is published in two increments, a
Northern Sample and a Southern Sample.  Both the SPOCS and GCS
catalogs provide age estimates and confidence intervals, from
which we construct two measures of a star's age.  The ``age'' samples
adopt the nominal stellar age while the ``eage'' samples use the lower
limit on the age estimate.  Column~3, ``magnetospheric subset,'' lists
the number of stars from each catalog passing the four selection criteria
of \S\ref{sec:catalog}.}
\tablerefs{(1)~\cite{gcg+03}; (2)~\cite{gcg+06}; (3)~\cite{tfsrfv07}; (4)~\cite{hna09}}
\end{deluxetable}

This search is ``blind'' in the sense that we have made no effort to
select stars already known to have planets, and even attempted to
select stars whose properties are such that they might not previously
have been searched for planets.  Nonetheless, within our samples are
small fractions of stars with known planets.  We have cross-correlated
our samples with the Extrasolar Planet Encyclopedia
\cite[version~July~1]{s09}.  The SPOCS-age sample has the highest
fraction of stars with known planets, 6\%, with the remaining samples
all having fractions below~5\%.  The relatively large fraction of
stars with known planets in the SPOCS-age sample is consistent with
the larger SPOCS catalog being drawn from a set of stars which are or
have been the focus of radial velocity searches.  The fraction of
stars in our samples that are known to have planets is smaller than
the estimated fraction of stars with planets \citep[19\%,][]{mbv+08},
consistent with the notion that current planetary searches continue to
be affected by selection effects.

\section{Magnetospheric Emission Searches}\label{sec:radio}

\subsection{Statistical Analysis}\label{sec:stack}

Our search is based on the VLA Low-frequency Survey
\citep[\hbox{VLSS}, ][]{clc+07}.  This survey imaged 95\% the sky
north of a declination limit $\delta > -30\arcdeg$ at a frequency
of~74~MHz (4~m wavelength) with a typical rms noise level
of~100~\mjybm.  Our focus on this survey is two-fold.  
First, the
frequency of this survey is within a factor of~2 of the cutoff
frequency of Jupiter ($\simeq 40$~MHz), so that it is plausible that
other Jovian-like planets might emit at~74~MHz.  
Second, it is an
electronically-available survey covering a large fraction of the sky,
so that many nearby stars are potentially accessible.  
There  are other surveys at
lower frequencies and which cover a significant fraction of the sky,
to which this approach might also be applied
\citep[e.g.,][]{r90,du_s90}
However, the VLSS has the advantage of having images that are readily
available in an electronic format combined with a high angular
resolution and sensitivity.

Each individual VLSS image was obtained by combining a series of
``snapshots'' acquired over a range of hour angles, with the time
sampling within a snapshot being 10~seconds.  A snapshot was typically
15--25~minutes in length, with snapshots separated typically by
about~1~hour.  For comparison, Jovian decametric emission observed
by the Nan{\c c}ay Decameter Array has been observed to have an average
duration of about~1~hr, with a range from about~0.5~hr to a few hours,
though these observations were probably dominated by the Io-controlled
component of Jovian decametric emission \citep{abmglr00}.  Although it
is too low in frequency to penetrate the \objectname{Earth's}
ionosphere, Saturian kilometric radiation observed by the
\textit{Cassini} spacecraft shows similar temporal characteristics
\citep{lzcpkg08}, but is not significantly affected by the presence of
a major satellite.  Assuming that extrasolar planetary magnetospheric
emissions are similar to those of \objectname{Jupiter} and
\objectname{Saturn}, if a planet was emitting during the course of a
VLSS observation, it was likely to have been emitting for at least the
duration of one snapshot, and potentially all of them.  Consequently,
there is a factor, potentially of order~1/3, in the luminosities that
we derive below that would account for the fact that a planet might
only be emitting for a fraction of the time that was used to acquire a
VLSS observation.  This factor is sufficiently small, relative to
other uncertainties, that we shall not incorporate it explicitly into
the analysis below. 

For each of our samples (Table~\ref{tab:catalog}), we downloaded small
images (``postage stamps'', 1\arcdeg\ in diameter) from the 
\anchor{http://lwa.nrl.navy.mil/VLSS/}{VLSS image server}.  Although
the formal declination limit of the VLSS is $-30\arcdeg$, the lowest
declination fields are the most incomplete and often have higher rms
noise levels.  Thus, we used an effective declination limit of~$-25\arcdeg$.

We aligned each postage stamp image so that the target star was in the
central pixel.  The beam (point spread function) of the VLSS was
80\arcsec, with an image pixel size of~25\arcsec.  The coordinates of
these stars are typically determined from the \textit{Hipparcos}
astrometric mission, epoch 1991.25 \citep{plk+97}; the VLSS
observations were conducted between~2001 and~2007, with the majority
of the observations conducted between 2003~September and 2005~April.
\cite{lf08} considered the possible astrometric uncertainties between
the \textit{Hipparcos} and VLSS frames, for the relatively high proper
motion star $\tau$~Boo (proper motion $\mu \simeq
0\farcs5$~yr${}^{-1}$).  They showed that the combination of
astrometric uncertainties, uncorrected ionospheric refraction within
the \hbox{VLSS}, and the proper motion of the star should have
produced an astrometric uncertainty in alignment of no more than
8\arcsec, a fraction of a pixel.  Thus, we are confident that
alignment to the central pixel in the postage stamp images is
sufficient.

We examined each of the images for any sources that might be confused
with a target star or that would be close enough to the location of a
target star to affect the stacking process.  As an example, the star
\objectname{HD~69582}, which appears in all of our samples, is
approximately 75\arcsec\ ($\approx 1$ beam) from the radio source
\objectname{PKS~0814$-$029}.  This radio source is generally
identified as a \hbox{QSO}, although no redshift has been measured.
The offset between the star and radio source is significant enough
that it is unlikely that the two are the same.
We return to the question
of individual stars below.
In addition, we found a small number of stars that are located close
to the boundaries of the \hbox{VLSS}, particularly near the southern
declination limit.  The noise level in the images for these stars was
much higher (and it is possible that the astrometry is not as
precise), so we excluded these.  The total number of stars
excluded on these considerations led to the sample sizes being
about~10 stars smaller than a straightforward application of our
initial selection criteria would suggest.

The number of stars~$N$ in each of our samples ranges from approximately
100 to several hundred.  Assuming that the noise in the VLSS images is
gaussian distributed, as generally expected for radio interferometric
images,\footnote{
In general, radio interferometric images of the sky are constructed
using a Fast Fourier Transform, with the flux density of the
visibility function at the spatial frequency origin taken to be
identically zero.  From Fourier Transform properties, this so-called
``zero-spacing'' visibility value is equivalent to the total flux
density within the field of view.  In the absence of a source, the
pixels in a thermal noise-limited image constructed in this manner
will have a zero-mean normal distribution.  The VLSS images that we
analyze were constructed using this standard procedure.}
we anticipate that the noise in a stacked image should be roughly
100$N^{-1/2}$~\mjybm, or about~5--10~\mjybm.  In our stacking
analysis, we combined the images in a weighted sense, weighting each
image's contribution to the final stacked image by its individual rms
noise level.  Using the NStars sample as an illustration (Table~\ref{tab:catalog}), and taking
the rms noise levels in the images into account, we expect that the
rms noise level in the stacked image should be 5.6~\mjybm.  The actual
rms noise level is 5.7~\mjybm, indicating that the assumption of
gaussian noise-dominated images for the VLSS is justified, at least to
the stacked image noise levels we have obtained here.
Table~\ref{tab:catalog} also presents the rms noise levels in the
stacked images.

In none of the stacked images do we detect statistically significant
emission.
In an area of approximately 1 beam in size, the strongest pixels range
from approximately 1.5$\sigma$ to~2.2$\sigma$.  Table~\ref{tab:search}
presents 3$\sigma$ limits on the average flux density of
magnetospheric emissions from planets orbiting the stars in our
samples from the various catalogs.  Table~\ref{tab:catalog} presents
the weighted average distance for the stars in the various samples,
with the weighted average distance for the $N$ stars in a sample
defined as 
\begin{equation}
\frac{1}{{\bar D}^2} \equiv \sum_{i=1}^N \frac{1}{D_i^2},
\label{eqn:dweight}
\end{equation}
where $D_i$ is the distance to the $i^{\mathrm{th}}$ star in the sample.
Using the weighted average distance, and assuming that the bandwidth
of planetary magnetospheric emissions is comparable to the observation
frequency, as it is in the case of Jupiter, we convert the flux
density upper limits to limits on average planetary luminosities.
Table~\ref{tab:search} also presents these planetary luminosity
limits.

\subsection{Individual Stars}\label{sec:stars}

Prior to stacking the images in each sample, we determined the peak
intensity around each star relative to the \emph{individual} image
noise level ($\approx 100\,\mjybm$).  Stars for which the peak
intensity exceeded 2.5$\sigma$ were then re-examined.

One motivation for performing this check is that the VLSS catalog of
sources was constructed using a threshold test, relative to the image
noise level, with a relatively high signal-to-noise threshold in order
to maintain a low probability for a false detection.  For the
\hbox{VLSS}, the signal-to-noise thresholds was 7$\sigma$.  Thus, it
is possible that there is stellar (or planetary) emission that would
not have been cataloged.  We found no stellar position for which radio
emission above~3$\sigma$ ($\approx 300\,\mjybm$) could be identified
unambiguously.  There were stellar positions with radio emission above
this level, but they could be explained by other features, such as a
sidelobe from another source.

As noted above, the positions of some stars were close to, if not
coincident with, radio sources, specifically the stars
\objectname{HD~38392}, \objectname{HD~49933}, \objectname{HD~79555},
\objectname{HD~143333}, and \objectname{HD~202575}.  These
stars have been detected by ROSAT \citep{hsv98,hssv99}, with X-ray
luminosities ranging from~$6 \times 10^{20}$~W to~$3 \times
10^{22}$~\hbox{W}.  The Benz-G\"udel relation predicts that the
centimeter-wavelength flux densities of these stars should be of order
0.1~mJy.  Scaling these flux densities to the VLSS (74~MHz, $\lambda =
4$~m), we expect no emission to be detectable, for any reasonable
radio spectral index.  We also examined the NVSS (1400~MHz, 20~cm)
near the location of each star.  The rms noise levels near these stars
are comparable, and a 3$\sigma$ upper limit to the radio emission on
any of these stars is 1.9~mJy, consistent with these stars not being
radio sources,  Finally, we are aware of targeted radio observations
of only one of these stars, \objectname{HD~143333}, which placed the
rather unconstraining limit of~2~Jy at~5~GHz \citep{bnt+92} on the
star or any associated planet.

\section{Discussion and Conclusions}\label{sec:conclude}

What do our limits imply about potential magnetospheric emissions from
planets orbiting any of the stars in our various samples?  As noted in
the discussion following equation~(\ref{eqn:law}), the planetary
luminosity can depend upon whether one is considering the kinetic
energy~$P_{\mathrm{sw,kin}}$ or magnetic energy~$P_{\mathrm{sw,mag}}$
carried by the solar wind.  The kinetic energy power depends upon the
stellar wind density~$n$ and velocity~$v$ as $P_{\mathrm{sw,kin}}
\propto nv^3$ while the magnetic energy power depends upon the stellar
wind velocity and magnetic field strength~$B$ as $P_{\mathrm{sw,mag}}
\propto vB^2$ \citep{ztrr01,gpkmmr07}.  The strengths of all of these
quantities both depend upon distance from the host star and are
expected to evolve with stellar age.

For the specific case of $\tau$~Boo, \cite{gpkmmr07} illustrated how
one can use a stellar wind model for a star with a known age and a
planet at a known orbital distance to estimate the strength of the
planetary radio emission.  A straightforward extension of their
approach could be applied to stars not yet known to be orbited by a
planet(s) and even samples such as those we have constructed here.
The additional relevant quantities that are needed would be the
distribution of planetary semi-major axes and, in the case of an
individual star, the distribution of its age estimate or, in the case
of a sample of stars, the distribution of their age estimates.
Combining these distributions, one could estimate the appropriate
scaling factor by which planetary magnetospheric emission would be
enhanced.  However, in \S\ref{sec:intro} we argued that the current
census of extragalactic planets is likely to be biased, particularly
with respect to those planets that might be most likely to be radio
emitting.  For that reason, we do not consider the distribution of
planetary semi-major axes to be well enough constrained to incorporate
it into our analysis, and we shall adopt a somewhat more simplified
approach below.

With respect to the stellar wind powers, \cite{gzs07} have synthesized
various observations and models to determine functional dependences
for stellar wind velocity, density, and magnetic field strength as a
function of stellar age (their equations~[15], [16], and~[23]--[24]),
for planets that are not too close to their host star.  Applying these
relations, and using the median age of the stars in the various
samples (Table~\ref{tab:catalog}), we determine a scaling factor,
relative to the current solar value at~1~\hbox{AU}, for each of these
quantities.  From the scaling factors for the individual quantities
($v$, $n$, and~$B$), the scaling factors for the kinetic energy and
magnetic energy powers are then determined.  Table~\ref{tab:search}
(Columns~4--8) presents these scaling factors.

These scaling factors are clearly only approximate and somewhat model
dependent.  The number of stars for which stellar wind parameters have
been determined is small.  Nonetheless, they serve as an indication of
the potential effect of stellar age.  We see that powers delivered
(at~1~AU) to potential planetary magnetospheres around these stars,
for most of these samples, may be enhanced by factors of~10--50 for
kinetic power and by factors of~5--10 for magnetic power.

The one potential exception is the GCS3-eage sample, for which much
larger stellar wind amplification powers appear possible.  These large
factors result from the relatively small median age for this sample
(0.7~Gyr, Table~\ref{tab:catalog}).  In turn, this small median age
likely reflects the relative lack of precision with which stellar ages
can be determined.  Many stars in the GCS3-eage sample have lower
limits to their stellar ages around~0.1~Gyr, because their age
estimates have large uncertainties (approaching 100\%).

These scaling factors for the stellar wind powers are for a fiducial
distance of~1~\hbox{AU}.  As noted above, there is likely to be a
strong star-planet distance dependence on the luminosity, but we do
not attempt to include a distribution of planetary semi-major axes.
Estimates of the star-planet distance dependence are that it is $d^x$,
with $x \lesssim 2$ \citep[e.g.,][]{fdz99,z07}, implying that Jupiter
would be about~25 times more luminuous were it at a distance of~1~AU
instead of its current 5.2~AU distance.  We therefore apply an
additional scaling factor of~25.

The final columns of Table~\ref{tab:search} present Jupiter's
luminosity scaled by these stellar wind kinetic energy and magnetic
energy factors, respectively, and the distance scaling factor of~25.
Even if a planet with luminosity comparable to Jupiter orbited every
one of our target stars, our sensitivity limits remain approximately
a factor of~10--100 above what would be needed to detect such
planets.  If the typical planet-star distance is larger than our
fiducial 1~AU value, the actual difference could be much larger.

We have not attempted to combine the stacked images from the various samples.
While the samples themselves are homogeneous, they obtain age
estimates from different methods.  Further, there are many stars that
are common to each sample, so that the stacked images are not independent.

There are a number of next-generation, low radio frequency instruments
under development.  Notable among these are the
\anchor{http://www.lofar.org/}{Low Frequency Array (LOFAR)} and the
\anchor{http://lwa.unm.edu/}{Long Wavelength Array (LWA)}.  If they
reach their design goals, both promise to provide rms noise levels
$\sigma \sim 3\,\mjybm$, at frequencies below~100~MHz, nearly two
orders of magnitude better than the 74-MHz VLA system which was used
to conduct the \hbox{VLSS}.  A similar statistical analysis applied to
future LOFAR or LWA observations may improve significantly upon the
limits presented here, or, ideally, detect extrasolar planetary emission.

Should either LOFAR or the LWA detect emission using a similar
statistical analysis, identifying the stacked emission as planetary
rather than stellar will be important.  In targeted observations of an
individual star or stars
\citep[e.g.,][]{bdl00,rzr04,gs07,lf08,scgjlb09}, an obvious
distinguishing factor would be whether the emission is modulated with
the planetary orbit.  For this statistical analysis, alternate aspects
of the stacked emission would have to be examined.  Both LOFAR and the
{LWA} are being designed to be broadband (over at least the 30--80~MHz
frequency range), thus the radio spectrum of the stacked emission
could be determined.  Further, any correlation between the strength of
stacked emission and the spectral type of the stars could be useful.

Building upon LOFAR and the \hbox{LWA} will be the
\anchor{http://www.skatelescope.org}{Square Kilometre Array (SKA)} and
the Lunar Radio Array (LRA).  While their designs will be influenced
by the work on \hbox{LOFAR}, \hbox{LWA}, and similar low radio
frequency interferometers, both the SKA and LRA currently anticipate
operating at frequencies that would be relevant for the detection of
planetary magnetospheres; in the case of the \hbox{SKA}, the design
goal for its lower operational frequency limit is 70~MHz, while, for
the \hbox{LRA}, frequencies $\nu \sim 50$~MHz are envisioned.  Both
would likely provide an order of magnitude sensitivity improvement
upon LOFAR and the \hbox{LWA}.

\acknowledgements
We thank N.~Kassim, S.~Kulkarni, and B.~Farrell for discussions which
spurred this analysis, A.~Cohen for discussions on the \hbox{VLSS},
B.~Erickson and B.~Slee for discussions about solar
decameter-wavelength emissions, and the referee for several comments
that improved the presentation of these results.  This research has
made use of the SIMBAD database, operated at \hbox{CDS}, Strasbourg,
France, and NASA's Astrophysics Data System.  Basic research in radio
astronomy at the NRL is supported by 6.1 Base funding.

\clearpage

\begin{deluxetable}{lccccccccc}
\rotate
\tablewidth{0pc}
\tabletypesize{\scriptsize}
\tablecaption{Magnetospheric Emission Limits\label{tab:search}}
\tablehead{%
                  & \colhead{Flux Density} & \colhead{Planetary}
	& \multicolumn{5}{c}{Stellar Wind Amplification Factors}
				    & \colhead{K.~E.~Scaled Jovian}
				    & \colhead{M.~E.~Scaled Jovian} \\
 \colhead{Sample} & \colhead{Upper Limit}
	& \colhead{Luminosity Limit\tablenotemark{a}}
	& \colhead{Velocity} & \colhead{Density} & \colhead{Magnetic Field}
	& \colhead{Kinetic Energy}
	& \colhead{Magnetic Energy} & \colhead{Luminosity\tablenotemark{a}}
	& \colhead{Luminosity\tablenotemark{a}} \\
		  & \colhead{(mJy)}        & \colhead{(erg~s${}^{-1}$)}
	& 		     &		         &
	&
	& 			    & \colhead{(erg~s${}^{-1}$)}
        & \colhead{(erg~s${}^{-1}$)}
}
\startdata
NStars     & 17 & $9.0 \times 10^{23}$ & 1.7 & 9.8 & 2.4 & 48 & 9.5 
	& $1.2 \times 10^{22}$ & $2.4 \times 10^{21}$ \\
SPOCS-age  & 33 & $1.1 \times 10^{24}$ & 1.4 & 4.9 & 1.8 & 15 & 4.8
	& $3.7 \times 10^{21}$ & $1.2 \times 10^{21}$ \\
SPOCS-eage & 28 & $5.1 \times 10^{23}$ & 1.6 & 8.6 & 2.2 & 38 & 8.3
	& $9.5 \times 10^{21}$ & $2.1 \times 10^{21}$ \\
GCS-age    & 18 & $7.3 \times 10^{23}$ & 1.6 & 6.7 & 2.0 & 25 & 6.5
	& $6.3 \times 10^{21}$ & $1.6 \times 10^{21}$ \\
GCS-eage   & 14 & $6.8 \times 10^{23}$ & 2.2 & 30  & 3.6 & 319 & 28
	& $8.0 \times 10^{22}$ & $7.1 \times 10^{21}$ \\
\enddata
\tablenotetext{a}{$1\,\mathrm{W} = 10^7\,\mathrm{erg}\,\mathrm{s}^{-1}$}
\tablecomments{%
The flux density upper limit and planetary luminosity limits are both
3$\sigma$.  The flux density limits are converted to planetary
luminosity limits using the weighted average distances for the samples
(Table~\ref{tab:catalog}) and assuming a 74~MHz bandwidth for the
planetary magnetospheric emissions.  We assume a fiducial planetary
distance of~1~AU and a planetary luminosity scaling with distance as
$d^2$, so that the kinetic and magnetic energy scaled Jovian
luminosities include an additional factor of~25; if the typical magnetospheric emitting planet is more distant
from (closer) its host star than 1~\hbox{AU}, these Jovian scaled
luminosities would have to be adjusted downward (upward).}
\end{deluxetable}

\end{document}